\documentclass[aps,prl,reprint,floatfix,superscriptaddress]{revtex4-1}
\usepackage[english]{babel}                                                           
\usepackage[utf8]{inputenc}                                                            
\usepackage[T1]{fontenc}
\usepackage{epsfig}
\usepackage{amsmath}
\usepackage{hyphenat}
\usepackage{hyperref}
\usepackage{placeins}
\usepackage[justification=raggedright]{caption}	
\usepackage[colorinlistoftodos]{todonotes}

\begin{document}

\title{Spin-liquid phase and order-by-disorder of classical Heisenberg spins on the swedenborgite lattice}
\author{Stefan Buhrandt}
\email[To whom correspondence should be addressed: ]{buhrandt@thp.uni-koeln.de}
%\affiliation{Institut für theoretische Physik, Universität zu Köln, Zülpicher Straße 77, 50937 Köln, Germany}
%\affiliation{Institute for Theoretical Physics, Utrecht University, Leuvenlaan 4, 3584 CE Utrecht, Netherlands}
\author{Lars Fritz}
\affiliation{Institut für theoretische Physik, Universität zu Köln, Zülpicher Straße 77, 50937 Köln, Germany}
\affiliation{Institute for Theoretical Physics and Center for Extreme Matter and Emergent Phenomena,
Utrecht University, Leuvenlaan 4, 3584 CE Utrecht, The Netherlands}

\begin{abstract}
Frustration refers to the inability to satisfy competing interactions simultaneously, often leading to a large number of degenerate ground states. This can suppress ordering tendencies, sometimes resulting in a spin liquid phase. An intrinsic effect lifting this degeneracy is entropic order-by-disorder. We present strong evidence that a classical nearest neighbor Heisenberg model on the swedenborgite lattice exhibits both an extended spin-liquid phase as well as entropic order-by-disorder choosing planar configurations after a first order transition. We argue that this observation renders magnetic insulators such as ${\rm{RBaCo}}_4{\rm{O}}_7$, where R denotes a rare earth atom, prime candidates for displaying spin liquid behavior and entropic order-by-disorder physics due to their large exchange constant.
\end{abstract}
\maketitle

%\section{Introduction}
{\it{Introduction}}
Frustration can exist whenever there is a set of interactions whose mutual satisfaction is excluded. A prime example are antiferromagnetically coupled spins on geometrically frustrated lattices. Geometric frustration, for instance present in two dimensions (2d) on the triangular or Kagom\'e lattices or in three dimensions (3d) on the pyrochlore lattice, does not automatically entail the existence of degenerate classical ground states, the hallmark of frustration. While the classical $120^\circ$-state of the 2d Heisenberg antiferromagnet on the triangular lattice is unique (up to global rotations and interchange of sublattices) there is a large number of degenerate ground states on the Kagom\'e lattice~\cite{Chalker1992}. The fluctuation between many degenerate classical ground states can lead to a suppression of ordering tendencies, resulting in a cooperative paramagnet or classical spin liquid~\cite{MoessnerChalker1998PRL,MoessnerChalker1998}. 
On a practical level, the ratio between the Curie-Weiss temperature $\Theta_{CW}$ (usually set by the coupling strength) and the ordering temperature $T_c$, {\it i.e.}, $f=\frac{|\Theta_{CW}|}{T_c}$ is a measure for frustration. Known systems with strong frustration have $f\approx 5-10$, thereby opening a window of spin liquid physics for $T_c<T<\Theta_{CW}$~\cite{Ramirez1994}. This degeneracy is fragile and usually small perturbations lift it: dipolar interactions plus disorder~\cite{Bramwell2001}, spin-lattice coupling~\cite{Yamashita2000}, and Dzyaloshinskii-Moriya interaction~\cite{Veillette2005} are known mechanisms. In quantum models, quantum fluctuations can lift the degeneracy at temperature $T=0$~\cite{Rastelli1987,Henley1989,Chubukov1992}. In classical systems, entropic order-by-disorder is a selection mechanism choosing one (or a subset) of the many degenerate classical ground states at small but finite $T$~\cite{Villain1980}. The decisive quantity is the free energy $F=E-T S$ which favors the state/s which best can exploit the low-energy configurations in its/their vicinity. A famous example of this mechanism is provided by the classical Kagom\'e antiferromagnet where planar ground state configurations are favored at low T~\cite{Chalker1992}, but other examples, also in higher dimensions, have been reported in the literature~\cite{Henley1989,Gvozdikova2005,Bergman2007}. For other 3d systems, e.g. Heisenberg spins on the pyrochlore lattice, it was shown to be completely absent, resulting in true classical spin liquids at $T=0$~\cite{Reimers1992,MoessnerChalker1998}. The factor responsible for presence or absence of order-by-disorder was identified to be the degree of degeneracy~\cite{MoessnerChalker1998}.  

\begin{figure}[h]
	\centering
	\includegraphics[width=\columnwidth]{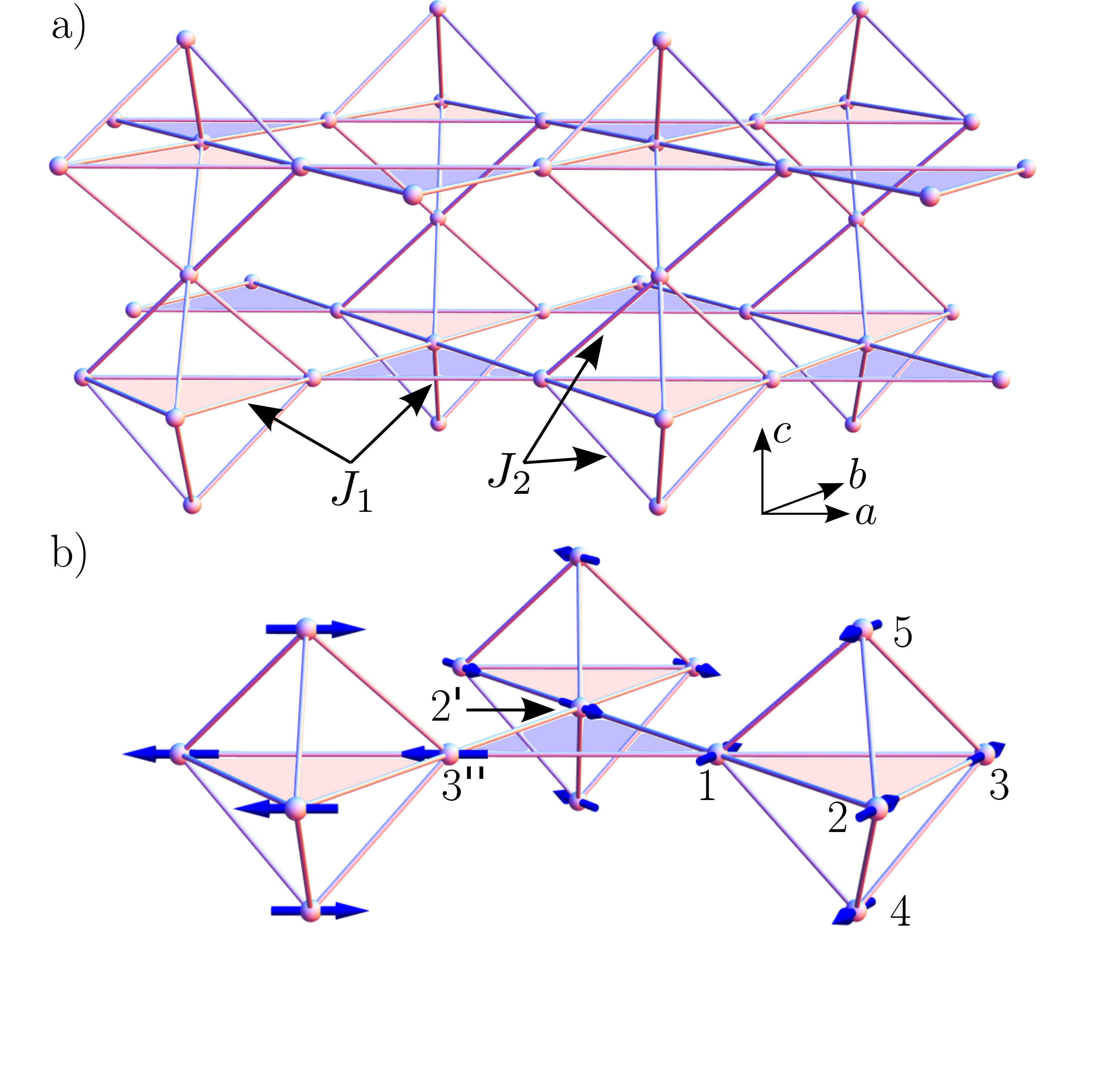}
	\caption{\textbf{a)} The lattice formed by the magnetic ions in swedenborgite compounds. \textbf{b)} Three bipyramids and a connecting intermediate triangle as elementary building blocks of the lattice. The intermediate blue triangle requires a $120^\circ$ configuration of the three spins around it. The spin configuration shown is the ground state configuration for $J_2/J_1\geq 3/2$.}
	\label{fig:lattice}
\end{figure}

Recently, a new class of geometrically frustrated structures based on cobalt oxides, ${\rm{RBaCo}}_4{\rm{O}}_7$, where R denotes a rare earth atom, emerged. The magnetic Co-atoms in this structure reside on the so-called swedenborgite lattice and at present there exist examples of systems that order magnetically and some that do not~\cite{Valldor2009,Chapon2006,Soda2006,Valldor2002,Manuel2009,Khalyavin2010,Schweika2007,Valldor2006}. In this paper we undertake an in-depth study of a minimal model for these compounds and show evidence that both classical spin liquid phases as well as entropic order-by-disorder are potentially existent in these systems.

%\section{Lattice structure and model}
{\it {Lattice structure and model: }}
The swedenborgite-lattice has a 3d hexagonal lattice structure with a non-trivial unit cell comprising eight atoms; there is a bipyramidal cluster of eight atoms stacked along $c$-direction and connected by intermediate triangles in the $ab$-plane (see Fig.~\ref{fig:lattice} a)). Alternatively, one can view it as an alternating stack of triangular and Kagom\'{e} layers (note that the structure repeats itself in c-direction). We consider a classical nearest neighbor Heisenberg model with only two distinct interactions: $J_1$ inside the Kagom\'{e} layers and $J_2$ between the Kagom\'{e} and triangular layers (Fig.~\ref{fig:lattice} a)), 
\begin{equation}
	H = J_1\sum_{ {\langle i,j \rangle \in \atop \text{same layer}}} \boldsymbol{S}_i \cdot \boldsymbol{S}_j + J_2\sum_{{\langle i,j \rangle \in \atop \text{diff. layer}}} \boldsymbol{S}_i \cdot \boldsymbol{S}_j \;,
	\label{eq:Hamiltonian}
\end{equation}
with classical $O(3)$ Heisenberg spins with $\|\boldsymbol{S}_i\|=1$ and antiferromagnetic interactions $J_1,J_2 > 0$ ($ \langle i,j \rangle$ refers to nearest neighbors). We rewrite Eq.~\eqref{eq:Hamiltonian} in terms of the elementary units, bipyramid and simple triangle (Fig.~\ref{fig:lattice} b) for the numbering of the spins), according to
\begin{eqnarray}
	H_{\text{bipyramid}} &= &\frac{J_1}{2} \left( \boldsymbol{S}_1 + \boldsymbol{S}_2 + \boldsymbol{S}_3 + \frac{J_2}{J_1}\left(\boldsymbol{S}_4 + \boldsymbol{S}_5\right) \right)^2 \nonumber \\
		&+&\frac{J_2^2}{2J_1} \left( \boldsymbol{S}_4 - \boldsymbol{S}_5 \right)^2 + \text{const.} \label{eq:Hamiltonianbipyramid}
\end{eqnarray}
and
\begin{eqnarray}		
		H_{\text{triangle}} = \frac{J_1}{2} (\boldsymbol{S}_1+\boldsymbol{S}_{2'}+\boldsymbol{S}_{3''})^2+\text{const.}\;.
	\label{eq:Hamiltoniantriangle}
\end{eqnarray}
While Eq.~\eqref{eq:Hamiltoniantriangle} is minimized by a planar $120^\circ$ configuration, within a bipyramid the best configuration satisfies
\begin{eqnarray}\label{eq:spinrule}
\boldsymbol{S}_4 - \boldsymbol{S}_5 &=&0 \;, \nonumber \\ \boldsymbol{S}_1 + \boldsymbol{S}_2 + \boldsymbol{S}_3 + \frac{J_2}{J_1}\left(\boldsymbol{S}_4 + \boldsymbol{S}_5\right)&=&0 \;. 
\end{eqnarray}
The first term aligns the triangular spins parallely, the second term wants the sum of the Kagom\'e spins antiparallel with the triangular spins~\cite{Manuel2009,Khalyavin2010}. 

If the system does not have more degrees of freedom than constraints on the ground state manifold (\textit{i.\,e.} if the ground state degeneracy is sufficiently small), it might show order-by-disorder. Otherwise, it remains disordered at all temperatures \cite{MoessnerChalker1998}. We have 8 spins per unit cell with two angle degrees of freedom each and consequently 16 degrees of freedom, whereas Eq.~\eqref{eq:spinrule} and $120^{\circ}$ order on the intermediate triangles impose together 16 constraints per unit cell. Consequently, we can expect an order-by-disorder transition for $J_2/J_1 < 3/2$ if soft modes are present.

\begin{figure}[h]
\includegraphics[width=\columnwidth]{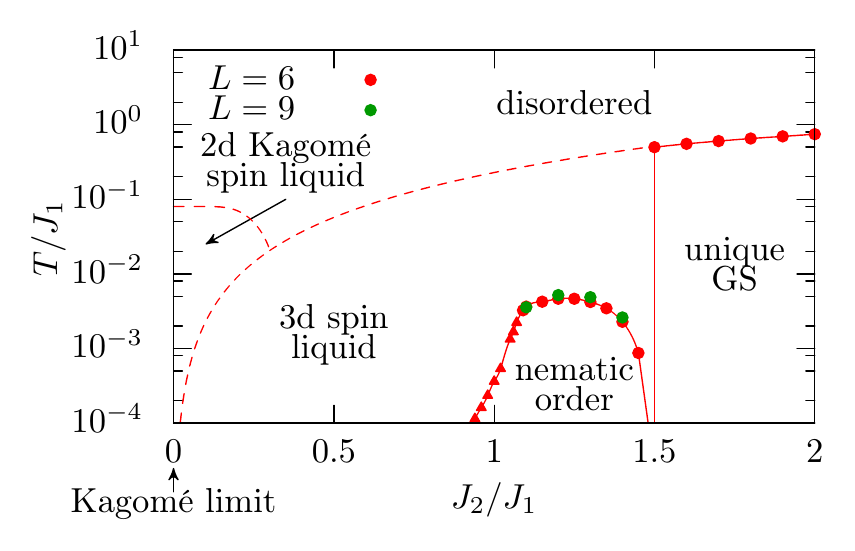}
\caption{Phase diagram: For $J_2/J_1\geq 3/2$ there is a unique ground state and the transition is second order. For $J_2/J_1<3/2$ the dashed line continues the second order transition but is a crossover line which is set by the scale $J_2^2/J_1$, separating a standard paramagnet from a 3d classical spin liquid. At low T there is a first order transition to a nematically ordered phase. The dots emerge from an analysis of the peak in the specific heat while the triangles are determined from the drop of the diffusivity of the replicas, see text for more explanations. This phase extends all the way to $J_2=0$ but is not shown due to the logarithmic temperature scale. The crossover line to the 2d spin liquid is determined along the lines of Zhitomirsky~\cite{Zhitomirsky2008}.}\label{fig:pd}
\end{figure}

The main result of our paper is the phase diagram of Eq.~\eqref{eq:Hamiltonian} shown in Fig.~\ref{fig:pd}. For $J_2/J_1\geq 3/2$, we find a transition to a magnetically ordered state, while for $0.3 \lesssim J_2/J_1<3/2$ there is a crossover from a conventional paramagnet to a cooperative paramagnet, a 3d classical spin liquid, which eventually develops nematic order in a first order transition at very low temperatures. In the weak interlayer coupling limit,  $0<J_2/J_1 \lesssim 0.3$, there is first a crossover from a classical paramagnet to a 2d spin liquid (consisting of decoupled Kagom\'{e} layers which individually form cooperative paramagnets~\cite{Zhitomirsky2008}), followed by a crossover to the aforementioned 3d spin liquid and eventually a first order transition to a state with nematic order. $J_2=0$ denotes the limit of decoupled Kagom\'e layers. In the following we discuss the phase diagram from an analytical and Monte Carlo point of view.

\begin{figure}[h]
	\centering
	\includegraphics[width=0.9\columnwidth]{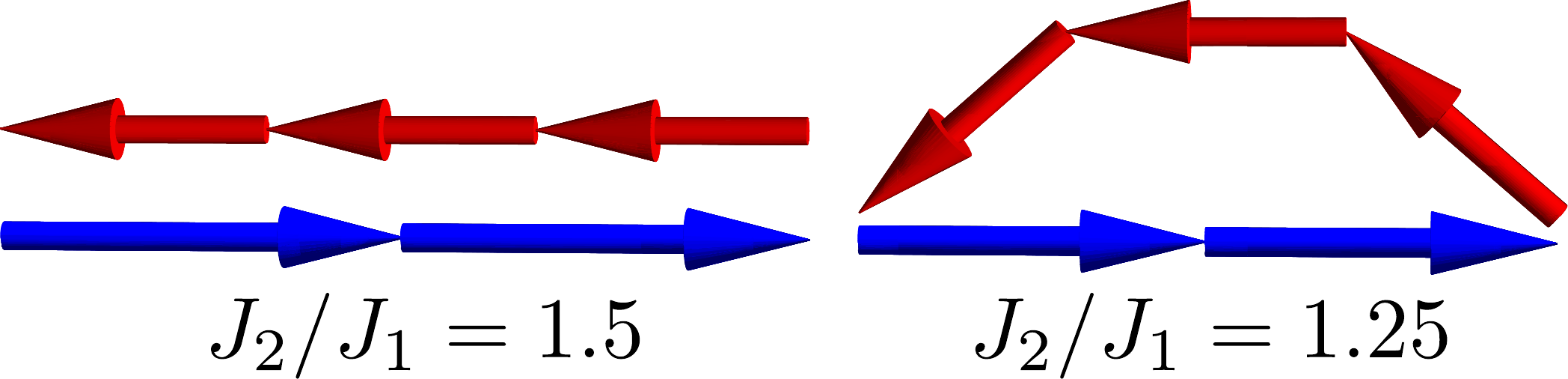}
	\caption{Ground state configurations of the Kagom\'{e} (red) and apical (blue) spins inside a bipyramid cluster for $J_2/J_1=1.5$ and $1.25$. The Kagom\'{e} spins have been rescaled by a factor $J_2/J_1$.}\label{fig:GS}
\end{figure}

\begin{figure}[h]
	\centering
	\includegraphics[width=0.9\columnwidth]{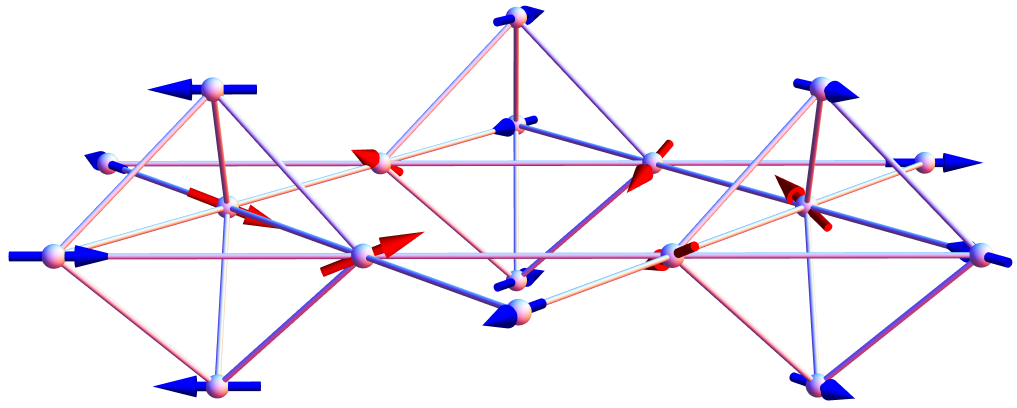}
	\caption{Local soft mode in coplanar ground state configurations for $J_2/J_1=3/2$: blue spins are idle while the red spins tilt out of the plane by $\pm \epsilon$ in an alternating fashion, leading to an energy cost $\omega\propto \epsilon^4$. }
	\label{fig:zeromode}
\end{figure}

%\subsection{Unique ground state for $J_2/J_1\geq 3/2$}
{\it{Unique ground state for $J_2/J_1\geq 3/2$:}}
With the apical spins being parallel, there exists a unique solution to Eq.~\eqref{eq:spinrule} for $J_2/J_1=3/2$ (up to global rotations), see left-hand-side of Fig.~\ref{fig:GS}. The Kagom\'e spins underneath/above the triangular spins are aligned antiparallely to the apical spins. To construct the ground state of the full system one has to stack the bipyramids in c-direction and ensure $120^\circ$-order at the intermediate triangles, Eq.~\eqref{eq:Hamiltoniantriangle}. This leads to a $120^\circ$ pattern of the triangular spins in the ab-plane. This configuration remains optimal also for $J_2/J_1>3/2$, even though the second condition in Eq.~\eqref{eq:spinrule} cannot be satisfied any more. The ground state is characterized by a broken spin rotation symmetry (plus breaking of translational symmetry) and is reached after a second order transition taking place at $T_{c}$. A classical spin wave analysis to harmonic level around the unique ground state reveals that two eigenmodes (localized around a hexagon in the Kagom\'e plane, see Fig.~\ref{fig:zeromode}) become soft for $J_2/J_1=3/2$ with a corresponding excitation energy $\omega \propto \epsilon^4$, where $\epsilon$ is the out-of-plane tilt. These soft modes leave a footprint in the specific heat: While a generic harmonic mode contributes a factor of $1/2$ per spin, a soft mode only contributes a factor $1/4$. A careful mode counting analysis along the lines of Chalker\cite{Chalker1992} predicts the specific heat to approach the value $15/16$ as $T \to 0$ for $J_2/J_1=3/2$ while it goes to one for $J_2/J_1>3/2$.

%\subsection{Spin-liquid phase and order-by-disorder for $0< J_2/J_1 <3/2$}
{\it{Spin-liquid phase and order-by-disorder for $0< J_2/J_1 <3/2$:}}
Eq.~\eqref{eq:spinrule} allows for many solutions, the same number as for a triangle in a magnetic field once the triangular spins are fixed~\cite{Kawamura1985,Chubukov1992,Seabra2011}. Joining the bipyramids via the intermediate triangles removes some of the degeneracy. On a global level this still allows for a large number of magnetically ordered ground states (but also disordered states). The fluctuations between these degenerate configurations possibly suppress a transition towards a (magnetically) ordered state altogether. However, we can find an argument favoring planar configurations characterized by a nematic order parameter, $Q^{\alpha \beta}=\frac{1}{N} \sum_i (S_i^\alpha S_i^\beta-\frac{1}{3} \delta_{\alpha ,\beta})$~\cite{Zhitomirsky2008}. Repeating the harmonic analysis around coplanar configurations reveals that the two soft modes with excitation energy $\omega \propto \epsilon^4$ discussed above for $J_2/J_1 = 3/2$ are still present for these states while they are absent for non-coplanar configurations. The soft nature of the coplanar states leads to entropic selection in a first order transition, as required by Landau theory due to the presence of a third order term $\propto {\rm {tr}}\left ( Q^3\right)$.

%\section{Monte Carlo analysis}
{\it {Monte Carlo analysis}}
We have performed classical Monte-Carlo simulations for the model~\eqref{eq:Hamiltonian} with periodic boundary conditions using parallel tempering for a broad temperature regime from $T/J_1=10^{-4}...10^1$. Previous attempts using simulated annealing~\cite{Khalyavin2010} were restricted to $T/J_1\geq 10^{-2}$ and missed important aspects of the physics (most importantly the order-by-disorder transition). A feedback algorithm~\cite{Katzgraber2006} selects temperatures for the PT simulation to maximize the current of replicas drifting through temperature space, thereby optimizing equilibration at all temperatures. 

\begin{figure}
	\includegraphics[width=\columnwidth]{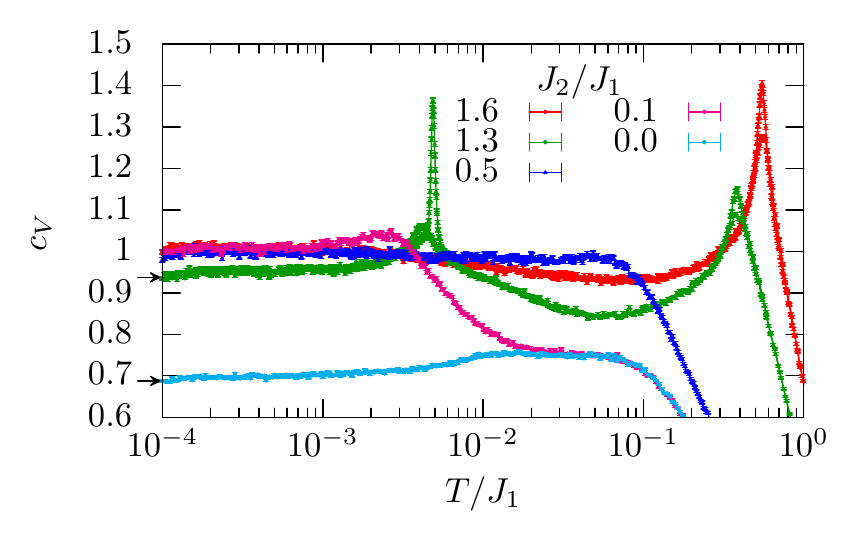}
	\caption{The specific heat $c_V$ for different ratios $J_2/J_1$ and $L=6 (N=1728)$ (as well as $L=9 (N=5832)$ for $J_2/J_1 = \{0.5,1.3,1.6\}$). In the limit $T \to 0$, the curve for $J_2=1.6$ goes to 1 while $J_2=1.3$ goes to $15/16$. For $J_2/J_1=0.1$ and $0.5$, the specific heat seemingly goes to 1 but the first order transition is pushed to very low temperatures. The arrows on the $c_V$-axis mark the values 15/16 and 11/16, the former being expected from our spin-wave analysis and the latter originating from rescaling the literature value known for the Kagom\'{e} lattice \cite{Chalker1992}, 11/12, by a factor 6/8 to account for the disordered triangular spins.}  
	\label{fig:cv}
\end{figure}

For $J_2/J_1 > 3/2$ we find a second order phase transition towards the symmetry broken phase with magnetic long-range order, indicated by a peak in the specific heat, c.\,f. Fig.~\ref{fig:cv} ($J_2/J_1=1.6$), which scales with system size. However, we were not able to determine the critical exponents to demonstrate that the transition is in the $O(3)$ universality class due to our restriction to too small system sizes with $N=8L^3, L=\{3,6,9\}$ (note that the magnetic unit, consisting of 24 spins, is very large).

For $J_2/J_1<3/2$, our analytical analysis has revealed a degenerate ground state manifold. The specific heat shows three different prototypical shapes in this region, see Fig.~\ref{fig:cv} for $J_2/J_1=1.3, 0.5$, and $0.1$. For $J_2/J_1=1.3$, there is a first peak at $T=\mathcal{O}(1)$ which marks the crossover from the conventional to the cooperative paramagnet. While the height of the peak seems to grow with system size in c-direction, it saturates in the ab-plane, incompatible with a second order transition. The correlation length in the ab-plane here is finite, while it is very long (but finite) in c-direction, c.\,f.~\cite{Valldor2011}. The corresponding spin structure factor does not show pronounced Bragg peaks concerning in-plane order and only quasi long-range correlations in c-direction survive upon increasing the system size. Upon decreasing $J_2$ to $J_2/J_1=0.5$ the correlations in ab-direction become more short ranged, the first peak in the specific heat vanishes continuously, and the specific heat looks increasingly like the one of a Heisenberg chain. Upon further decreasing $J_2$ down to $J_2=0.1$, we find that the system first behaves like its 2d Kagom\'e counterpart (compare curve with $J_2=0$), i.\,e., the Kagom\'e planes individually form cooperative paramagnets~\cite{Zhitomirsky2008}, which is why we call the regime 2d Kagom\'e spin liquid. Upon further lowering temperature the interlayer coupling $J_2$ becomes relevant and a crossover to the 3d spin liquid is observed. This suggests an interpretation of the 3d classical spin liquid as a system with short-range correlations in the ab-plane and quasi-long-range order in c-direction. This reasoning is further backed up by the position of the characteristic hump in the specific heat scaling like $J_2^2/J_1$, which is the effective ferromagnetic coupling of the triangular spins in c-direction, c.f. Eq.~\eqref{eq:Hamiltonianbipyramid}.

\begin{figure}
	\includegraphics[width=\columnwidth]{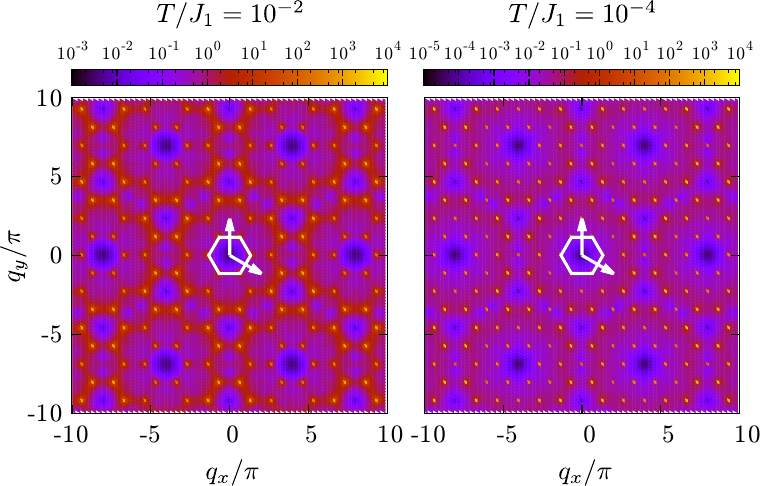}
	\caption{Structure factors in the $(q_x,q_y,q_z=0)$ plane for $J_2/J_1=1.3$ above ($T/J_1=10^{-2}$, left) and below ($T/J_1=10^{-4}$, right) the transition to the nematic state. Additional satellite peaks at the reciprocal lattice vectors $\boldsymbol{k}_1 = 2\pi(1,-1/\sqrt{3},0)^T,\boldsymbol{k_2}=2\pi(0,2/\sqrt{3},0)^T$ (white arrows) and points related by symmetry appear in the nematic phase.}  
	\label{fig:sf}
\end{figure}

For low temperatures, there is another sharp peak in the specific heat curve for $J_2/J_1=1.3$ corresponding to the aforementioned order-by-disorder transition. A jump in the nematic order parameter clearly indicates that the nature of this phase transition is first order. This transition becomes stronger with decreasing $J_2/J_1$ and could not be equilibrated for $J_2/J_1 \lesssim 1.1$. Nevertheless, it is possible to monitor the diffusivity of the different replicas through temperature space along the lines of \cite{Katzgraber2006} and determine the transition temperature from its pronounced drop at the phase transition, see the triangles in Fig.\ref{fig:pd}. After the transition, the specific heat takes the value $\lim_{T \to 0} c_V = 15/16$ (see curve for $J_2=1.3$), as predicted by the mode counting analysis (see Fig.~\ref{fig:cv}). The nematic transition is also accompanied by the appearance of additional satellite peaks at the reciprocal lattice vectors $\boldsymbol{k}_1 = 2\pi(1,-1/\sqrt{3},0)^T,\boldsymbol{k_2}=2\pi(0,2/\sqrt{3},0)^T$ and wave-vectors related by symmetry, see Fig.\ref{fig:sf}. Similar behavior is found on the Kagom\'e lattice \cite{Zhitomirsky2008}

From our numerical analysis the phase boundary between the magnetically ordered phase for $J_2/J_1 \geq 3/2$ and the 3d spin liquid phase for $J_2/J_1<3/2$, see Fig.~\ref{fig:pd}, cannot distinguish between an actual vertical phase boundary and a steep slope towards $J_2/J_1>3/2$. For entropic reasons we suspect the latter scenario.

%\section{Conclusion and Outlook}
{\it{Conclusion and Outlook:}}
Using a combination of analytical considerations and classical Monte Carlo we have shown that spin systems on the swedenborgite lattice offer an exciting playground in which one can hope to encounter both spin liquid phases as well as entropic order-by-disorder. To show this we have determined the phase diagram of a nearest-neighbor antiferromagnetic Heisenberg model with two distinct exchange interactions, see Fig.~\ref{fig:pd}. While the magnetically ordered ground state is unique for $J_2/J_1 \geq 3/2$, it becomes degenerate for $J_2/J_1 < 3/2$. For $J_2/J_1 \geq 3/2$ there is a second order transition to the magnetically ordered state. For $J_2/J_1<3/2$ this transition ceases to exist and the system crosses over to a regime where it fluctuates between many degenerate spin configurations compatible with Eq.~\eqref{eq:spinrule} and the sum rule of the intercalated triangles. For very low temperatures, there is a first order transitions towards a phase with nematic order, in which among the many degenerate solutions the planar ones are favored. These coplanar ground state configurations feature soft-modes with excitation energy $\omega(\epsilon) \propto \epsilon^4$ which drive the entropic order-by-disorder mechanism. From our analysis we cannot exclude an entropic selection of a subset of the planar states at even lower temperatures~\cite{Chern2013}.  
With the exchange coupling in these systems being very high, {\it i.e.}, $J \approx 2000$~K~\cite{Valldor2006}, the experimental observation of a classical three dimensional spin liquid phase as well as the elusive order-by-disorder mechanism seem conceivable if systematic studies of compounds with different chemical compositions are performed. The ratio $J_2/J_1$ can be altered experimentally by appropriate chemical substitutions of the ions on the triangular layers. Actual compounds show a large spread in the realized ratio $J_2/J_1$ from $0\lesssim J_2/J_1 \ll 1$~\cite{Schweika2007,Stewart2011} in Y$_{0.5}$Ca$_{0.5}$BaCo$_4$O$_7$, $0.5$ in YBaCo$_3$FeO$_7$~\cite{Valldor2011}, $1$ in YBaCo$_4$O$_7$~\cite{Manuel2009} and $>3/2$ in CaBaCo$_2$Fe$_2$O$_7$~\cite{Reim2011}.   

%\section{Acknowledgement}
{\it {Acknowledgement:}}
We acknowledge discussions with J. Chalker, R. Moessner, K. Penc, N. Shannon, S. Trebst, M. Valldor, M. Vojta, and above all J. Reim and W. Schweika. This work was supported by the Deutsche Forschungsgemeinschaft within the Emmy-Noether program through Grant No. FR 2627/3-1 (S.B. and L.F.) and the Bonn-Cologne Graduate School for Physics and Astronomy (S.B.). This work is also part of the D-ITP consortium, a program of the Netherlands Organisation for Scientific Research (NWO) that is funded by the Dutch Ministry of Education, Culture and Science (OCW). Simulations were performed on the CHEOPS cluster at RRZK Köln.

\bibliography{references}

\end{document}